\documentclass[twocolumn,showpacs,preprintnumbers,amsmath,amssymb]{revtex4}
\usepackage{graphicx}% Include figure files

\usepackage{dcolumn}% Align table columns on decimal point
\usepackage{bm}% bold math

\newcommand{\be}{\begin{equation}}
\newcommand{\ee}{\end{equation}}
\newcommand{\bea}{\begin{eqnarray}}
\newcommand{\eea}{\end{eqnarray}}

\newcommand{\Hrel}{\hat{H}_{\mathrm{rel}}}

\newcommand{\Vhf}{\hat{V}_{\mathrm{hf}}}
\newcommand{\Vd}{\hat{V}_{\mathrm{d}}}

\newcommand{\Psil}{| \Psi^l \rangle}
\newcommand{\psils}{| \psi_S^l \rangle}
\newcommand{\alphastate}{| \alpha \rangle}

\newcommand{\aB}{a_{\mathrm{B}}}
\newcommand{\Spn}{\mathbf{S}}

\newcommand{\abm}{\mathrm{ABM}}

\newcommand{\br}{B_0}
\newcommand{\brexp}{B_0^{\mathrm{exp}}}

\newcommand{\DB}{\Delta B}
\newcommand{\DBexp}{\Delta B_{\mathrm{exp}}}

\newcommand{\gel}{g_e}
\newcommand{\gnu}{g_n}

\newcommand{\ET}{E_\mathrm{triplet}}
\newcommand{\ES}{E_\mathrm{singlet}}

\begin{document}

% \preprint{APS/??}

\title{Feshbach resonances in ultracold $^{85}$Rb--$^{87}$Rb and $^{6}$Li--$^{87}$Rb mixtures}
\author{Z. Li$^2$, S. Singh$^1$, T. V. Tscherbul$^2$, and K. W. Madison$^1$}
\affiliation{
$^{1}$Department of Physics and Astronomy, University of British Columbia, Vancouver, Canada V6T 1Z1\\
$^{2}$Department of Chemistry, University of British Columbia,
Vancouver, Canada, V6T 1Z1}

\date{\today}

\begin{abstract}
We present an analysis of experimentally accessible magnetic Feshbach resonances in ultracold heteronuclear $^{85}$Rb--$^{87}$Rb and $^{6}$Li--$^{87}$Rb mixtures.  Using recent experimental measurements of the triplet scattering lengths for $^{6}$Li--$^{87}$Rb and $^{7}$Li--$^{87}$Rb mixtures and Feshbach resonances for one combination of atomic states, we create model potential curves and fine tune them to reproduce the measured resonances and to predict the location of several experimentally relevant resonances in Li--Rb collisions.  To model $^{85}$Rb--$^{87}$Rb collisions, we use accurate Rb$_2$ potentials obtained previously from the analysis of experiments on $^{87}$Rb--$^{87}$Rb collisions. We find resonances that occur at very low magnetic fields, below 10 G, which may be useful for entanglement generation in optical lattices or atom chip magnetic traps.
\end{abstract}

\maketitle
\section{Introduction}
The field of ultracold atomic gases has been revolutionized by the discovery and use of magnetically tunable Feshbach resonances \cite{Tiesinga93,Weiner99,Jones06}.  These resonances occur when the energy of a quasibound molecular state (``closed" channel) is tuned into degeneracy with the energy of a colliding atomic pair in an ``open" channel by an externally applied magnetic field.  Near a resonance, the coupling between the states strongly affects both the elastic and inelastic collision cross sections of the colliding atoms.  First observed at low temperatures in homonuclear collisions of Na atoms \cite{Inouye98}, Feshbach resonances provide a way to tune the microscopic interactions and to coherently (reversibly) form ultracold molecules \cite{Donley02}.  These features have been exploited for the study of a wide variety of phenomena in ultracold atomic gases including the controlled collapse of a Bose-Einstein condensate (BEC) \cite{Claussen02}, bright solitons \cite{Strecker02, Khaykovich02}, and BCS superfluid pairing of a Fermi gas \cite{Chin04, Zwierlein04, Regal04}.

Mixtures of different ultracold atoms are interesting objects of study as they may exhibit scattering properties different from those of indistinguishable species. For
example, broken exchange symmetry allows for s-wave scattering of different fermions in the same hyperfine state. In addition, the Feshbach spectrum of heteronuclear molecules is more complex because the hyperfine
constants are unequal for the two atoms and there are four rather
than three zero-field collision asymptotes\cite{Bhattacharya04}.
Heteronuclear Feshbach resonances have recently been observed in 
$^{6}$Li--$^{23}$Na \cite{Stan04}, $^{40}$K--$^{87}$Rb \cite{Inouye04}, $^{85}$Rb--$^{87}$Rb \cite{Papp06}, $^{6}$Li--$^{87}$Rb \cite{Deh08}, and $^{6}$Li--$^{40}$K collisions \cite{Wille08}.
These resonances provide a mechanism to tune interspecies interactions and may enable the study of boson-mediated cooper pairing \cite{Heiselberg00} and supersolid order \cite{Buchler03} predicted to occur in Bose-Fermi mixtures.
Feshbach resonances in heteronuclear systems can be used to produce ultracold polar molecules \cite{Sage05}.  The dynamics of ultracold polar molecules is determined by tunable and anisotropic electric dipole-dipole interactions, and the creation of ultracold polar molecules is predicted to allow for major advance in several different fields of physics and chemistry.

In this paper, we analyze collisions in ultracold $^{85}$Rb--$^{87}$Rb 
and
$^{6}$Li--$^{87}$Rb mixtures.  Recently, Tiesinga combined the results of spectroscopic and scattering experiments on $^{87}$Rb ensembles to produce a new set of accurate Rb$_2$ interaction potentials of singlet and triplet symmetry \cite{Tiesinga08}. With the new potentials, Tiesinga was able to reproduce the experimentally observed Feshbach resonances in ultracold collisions of $^{87}$Rb atoms with a high level of accuracy. Here, we extend this work to heteronuclear rubidium dimers.
We calculate the magnetic field dependence of $^{85}$Rb--$^{87}$Rb elastic scattering cross sections
using the new interaction potentials and find good agreement with the recent experiment of
Papp and Wieman \cite{Papp06} and the theoretical analysis of Burke {\it et al}.\ \cite{Burke98}.  We also study ultracold scattering of Rb isotopes in hyperfine states other than the fully spin-stretched state $|2,-2\rangle_{85}\otimes |1,-1\rangle_{87}$ and find resonances that occur at low magnetic fields, below 10 G.  For the Li--Rb system, we create model potentials combining the results from recent \textit{ab initio} calculations for the LiRb dimer \cite{Aymar05,Derevianko01} and then fine tune them to reproduce the experimentally measured resonances.  For this purpose we utilize an asymptotic bound state model \cite{Moerdijk95, Stan04,Wille08} to determine the energies of the last bound states of the triplet and singlet potentials consistent with the experimentally observed location of two Feshbach resonances.  Guided by the results of the model analysis, we generate the corresponding potentials and compute both the location and widths of the Feshbach resonances using a full quantum scattering calculation.
This process allows us to systematically converge to the optimal singlet and triplet potentials for the Li--Rb dimer.  These potentials fully characterize the scattering properties in any combination of atomic spin states and allow us to predict both the triplet scattering length and the location of several experimentally accessible resonances for this mixture.

\subsection{Calculation method}
The Hamiltonian for the collision of two alkali metal atoms is
$H =
\Hrel + \Vhf + \Vd + \hat{V}_B$
where the first term accounts for the relative motion of the atoms, $\Vhf $ models the hyperfine interactions, and $\hat{V}_B$ models the interaction of the collision complex with the external magnetic field.  We have verified that the magnetic dipole interaction, $\Vd$, has no effect on the observables described in this paper so we neglect it.
\be
\Hrel = -\frac{1}{2\mu R}\frac{\partial^2}{\partial
R^2}R + \frac{\hat{l}^2}{2 \mu R^2}+ \hat{V}(R),
\label{eq:Hrel} 
\ee
where $\mu$ is the reduced mass of the atoms, $R$ is the interatomic
distance and $\hat{l}$ is the rotational angular momentum of the
collision complex. 
The total scattering
wave function is expanded in a fully uncoupled, space-fixed basis set:
\be
|\Psi\rangle = \frac{1}{R} \sum_{\alpha, l, M_l} F_{\alpha, l, M_l}(R) | l M_l\rangle  |\alpha\rangle \\
\ee
where $F_{\alpha, l, M_l}(R) | l M_l\rangle$ 
are the radial basis states and
$|\alpha\rangle = |I_a M_{I_a}\rangle |S_a M_{S_a}\rangle |I_b M_{I_b}\rangle |S_b M_{S_b}\rangle$ are the atomic spin states.
The electron and nuclear spins for atom $a$ ($b$) are denoted by $\Spn_a$ ($\Spn_b$) and $\mathbf{I}_a$ ($\mathbf{I}_b$).
The total electron spin for the complex is the vector sum of the electron spins of atoms $a$
and $b$, $\Spn = \Spn_a + \Spn_b$.
$\hat{V}(R) = \sum_{S} V_S(R) P_S$ describes the inter-atomic interaction potential
where 
$P_S = \sum_{M_S} |S M_S\rangle \langle S M_S|$
is the projection operator onto the singlet $S=0$
or triplet $S=1$ electron spin configuration of the molecule and
$V_S(R)$ is the corresponding interatomic potential for the singlet $X^1\Sigma$ or triplet $a^3\Sigma$ state of LiRb.
The interaction of the
collision complex with the external magnetic field
is given by
\begin{equation}
\hat{V}_B = \gel (\Spn_a + \Spn_b) \cdot \mathbf{B}
- (\gnu^{(a)} \mathbf{I}_{a} +  \gnu^{(b)} \mathbf{I}_{b}) \cdot \mathbf{B}
\end{equation}
where $\gel$ and $\gnu$ are the electron and nuclear gyromagnetic ratios.
The hyperfine interaction is described by
\begin{equation}
\Vhf = \gamma_a \mathbf{I}_a \cdot \mathbf{S}_a + \gamma_b
\mathbf{I}_b \cdot \mathbf{S}_b
\end{equation}
where $\gamma_a$ and $\gamma_b$ are the hyperfine interaction
constants for atoms $a$ and $b$.

The matrix elements of the Hamiltonian in the fully uncoupled basis are determined as described in Ref.\ \cite{Li07}, and the resulting coupled-channel equations are propagated into the asymptotic region where the standard asymptotic boundary
conditions \cite{Krems04b} are applied to extract the scattering matrix elements. The integral cross sections
and the scattering lengths are computed from the $S$-matrix elements as described in Ref.\ \cite{Hutson07}.

Using this procedure, we calculate the elastic scattering cross sections as functions of the magnetic field at a fixed collision energy.  To ensure the convergence of the calculations, we employ a dense propagation grid from $2.0$ to $800 \; \aB$ ($\aB = 0.0529177$~nm) with a step of $0.005 \; \aB$.
The $s$-wave scattering length near a Feshbach resonance has the form \cite{Hutson07}
\begin{equation}\label{eq:ResForm}
a(B) = a_\text{bg}\left( 1 - \frac{\DB}{\br-B}  \right),
\end{equation}
where $\br$ is the position and $\DB$ is the width of the resonance. The background scattering
length $a_\text{bg}$ does not depend on the magnetic field.  For each resonance, we extract the parameters $\br$ and $\DB$ from the magnetic field
dependence of the scattering length using Eq.\ \ref{eq:ResForm}. In the case of the 
$^{85}$Rb--$^{87}$Rb resonances, we have verified our results for $\br$ 
by analyzing the derivative of the eigenphase sum \cite{Krems04a}. The resonance positions
obtained in this way are found to be in excellent agreement with the results of Eq.\ \ref{eq:ResForm}. 

\section{Low field resonances in $^{85}$Rb--$^{87}$Rb mixtures}
%\subsection{Motivation}

In 2006, Papp and Wieman created heteronuclear molecules in a mixture of $^{85}$Rb and $^{87}$Rb
atoms by a linear ramp of the magnetic field and by resonant-field modulation \cite{Papp06}. For this purpose they used
two Feshbach resonances in the $|2,-2\rangle_{85}\otimes |1,-1\rangle_{87}$ magnetic hyperfine state
(where $|f, m_f\rangle$ is the usual notation for the atomic hyperfine states).
Burke {\it et al}.\ considered magnetic Feshbach resonances in the 
weak magnetic field seeking states of $^{85}$Rb--$^{87}$Rb
and suggested optimal conditions for sympathetic cooling of $^{85}$Rb atoms by elastic collisions
with evaporatively cooled $^{87}$Rb \cite{Burke98}.  Using accurate Rb$_2$ 
potentials of singlet and triplet symmetry \cite{Tiesinga08}, we analyze the magnetic field dependence of $^{85}$Rb--$^{87}$Rb elastic scattering cross sections.  Our primary motivation for this work was to search for low field Feshbach resonances which would be candidates for the realization of RF induced Feshbach resonances using large RF or microwave fields generated near the surface of an atom chip \cite{Hofferberth06}.

\subsection{Results}

\begin{figure}[t!]
	\centering
	\includegraphics[width=8 cm]{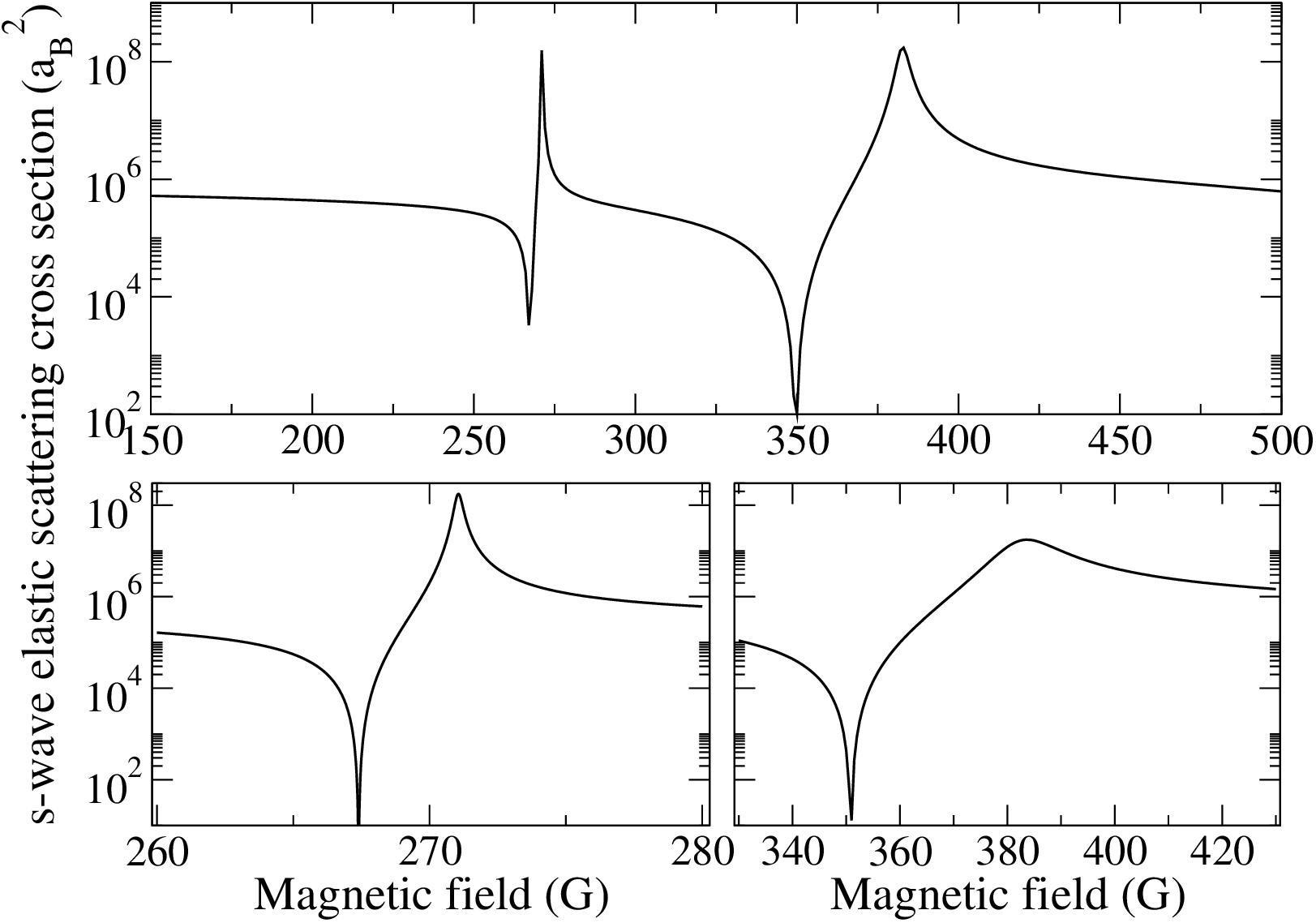}
	\renewcommand{\figurename}{Figure}
	\caption{Elastic scattering cross sections for $s$-wave collisions of $^{85}$Rb and $^{87}$Rb atoms
	in the lowest weak magnetic field seeking hyperfine state $|2,-2\rangle_{85}\otimes |1,-1\rangle_{87}$ versus
	the applied magnetic field (upper panel). The collision energy is 144 nK (see text). The lower panels
	show more details of the resonances.}
	\label{fig:8587crosssections}
\end{figure}

Figure \ref{fig:8587crosssections} shows the $s$-wave elastic scattering cross sections for $^{85}$Rb--$^{87}$Rb collisions in the weak magnetic field seeking state $|2,-2\rangle_{85}\otimes |1,-1\rangle_{87}$. Two resonances are immediately apparent
from this graph: the one at lower energy (resonance I) and a broader peak around 380 G (resonance II).
These resonances are of significant interest for sympathetic cooling
of $^{85}$Rb by collisions with ultracold $^{87}$Rb \cite{Burke98,Papp06}, and it is important to know
their paramaters with high accuracy. Table \ref{tab:I} compares our calculated resonance positions and widths
with recent experimental results of Papp and Wieman \cite{Papp06}. The agreement is generally good, although
the calculated position of resonance I (271.05 G) is shifted to higher magnetic fields by 5.6 G with
respect to the experimental value \cite{Papp06}.
The calculated width of resonance I is $\DB= 3.66$~G, consistent with the experimentally measured width, $\DBexp=5.8\pm 0.4$~G, of the trap loss feature \cite{Papp06}.  Although they are related, the width of the trap loss feature,  $\DBexp$, is not equivalent to the width 
of the scattering length singularity $\DB$ \cite{Wille08}.  Table \ref{tab:I} also shows the results obtained with a different set of interaction potentials by Burke {\it et al}.\ \cite{Burke98}. The lower bound
of their estimate almost coincides with the experimental prediction, whereas the upper bound is only 2 G away
from our result of 271.05 G.

\begin{table}
\caption{Positions and widths of magnetic Feshbach resonances in $^{85}$Rb--$^{87}$Rb collisions
for different initial atomic hyperfine states.
All atomic states considered are from the lowest hyperfine manifold
($f=1$ for $^{87}$Rb, $f=2$ for $^{85}$Rb). The numbers in parentheses are taken from calculations
of Burke {\it et al}.\ \cite{Burke98}.}
\begin{ruledtabular}
\begin{tabular}{cccccc}
 \multicolumn{2}{c}{atomic states} &  \multicolumn{3}{c}{Theory} & Experiment \\
$|f, m_{f}\rangle_{85}$ & $|f, m_{f}\rangle_{87}$ &  $\br$ & $\DB$ & $\br$\cite{Burke98} & $\brexp$ \\
 & & (G) & (G) & (G) & (G) \\
\hline
$|2,-2\rangle$  &  $|1,-1\rangle$  &  271.05 & 3.66 & $(267\pm 2)$  & $265.44\pm 0.15$ \\
                        &                          &  382.70 & 33.2 &  $(356\pm 3)$  & $372.4\pm 1.3$ \\
$|2,0\rangle$  &  $|1,-1 \rangle$  &  3.35 & 2.02   &                         & $\cdots$             \\
$|2,-1\rangle$  &  $|1,-1\rangle$  &  5.51  & 0.84  &                        & $\cdots$             \\
$|2,1\rangle$  &  $|1,-1 \rangle$  &  3.55    & 1.36  &                       & $\cdots$             \\
\end{tabular}
\end{ruledtabular}
\label{tab:I}
\end{table}

The second resonance in Fig.\ \ref{fig:8587crosssections} is a broad resonance centered at 382.7 G with a calculated width of 33.2 G.
Table \ref{tab:I} shows that our calculations overestimate the resonance position by 10 G. On the other hand, Burke {\it et al}.\ \cite{Burke98} predicted resonance II to occur at $356\pm3$ G which is 26 G lower than
the observed value \cite{Papp06}. Taking into account that the experimental uncertainty in the position of the broad resonance II may be up to a few Gauss (it is not given in Ref.\ \cite{Papp06}), the agreement between our calculations and experiment can be considered very good.

\begin{figure}[t!]
	\centering
	\includegraphics[width=8cm]{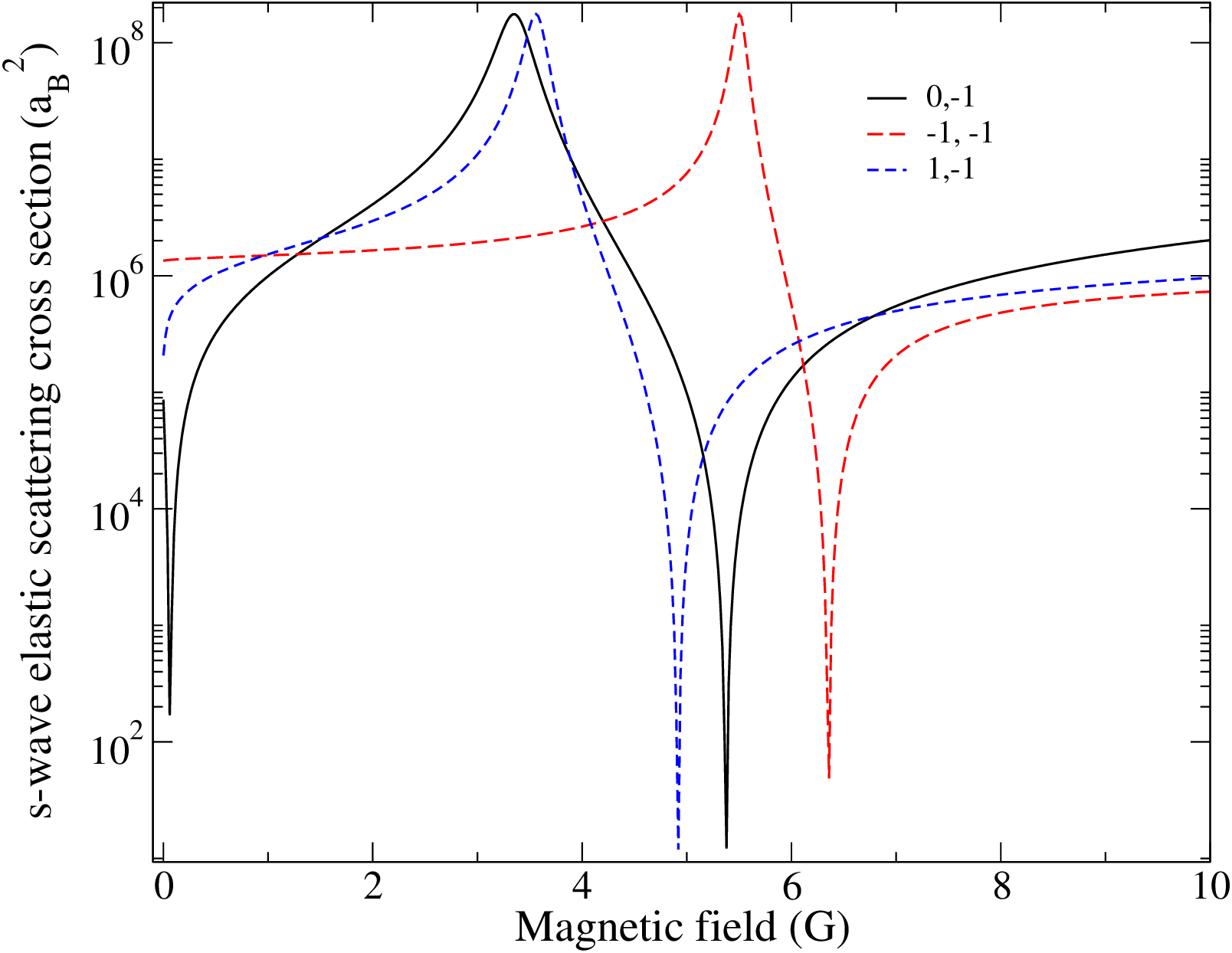}
	\renewcommand{\figurename}{Figure}
	\caption{Magnetic field dependence of $s$-wave elastic scattering cross sections for $^{85}$Rb--$^{87}$Rb
	in different hyperfine states. Full line -- $|2,0\rangle_{85}\otimes |1,-1\rangle_{87}$, long dashed line
	-- $|2,-1\rangle_{85}\otimes |1,-1\rangle_{87}$, short dashed line -- $|2,1\rangle_{85}\otimes |1,-1\rangle_{87}$.
	The collision energy is 144 nK.}
	\label{fig:LowFieldResonances}
\end{figure}

The resonances shown in Fig.\ \ref{fig:8587crosssections} occur when the atoms are in weak field-seeking, magnetically trappable states.  Optical traps, relying on the AC Stark shift, can provide confinement for arbitrary spin mixtures, and Feshbach resonances in high magnetic field-seeking states of $^{87}$Rb were recently observed at magnetic fields as small as a few Gauss \cite{Widera04,Erhard04}.

Figure \ref{fig:LowFieldResonances}  shows the $s$-wave Feshbach resonances in collisions of $^{87}$Rb in the $|1,-1\rangle$ state with $^{85}$Rb in 
three different hyperfine states: $|2,0\rangle$, $|2,-1\rangle$, and $|2,1\rangle$.
All the resonances occur at magnetic fields below 10 G, and have widths of order 1-2 G (see Table \ref{tab:I}).
They are similar to the 9 G resonance found in a gas of pure $^{87}$Rb \cite{Erhard04}.
We verfied that no other Feshbach resonances of similar widths occur at magnetic fields below 110~G.  The resonances shown in Fig.\ \ref{fig:LowFieldResonances} can be used to generate two-body spin entanglement in optical lattices \cite{Widera04}, and to create heteronuclear molecules from atoms in different hyperfine states.
Finally, since these resonances occur at very low magnetic fields, they are technically much easier to access and they are possible candidates for the realization of RF induced Feshbach resonances using large RF or microwave fields generated near the surface of an atom chip
 \cite{Hofferberth06}.

\section{Feshbach Resonances in $^{6}$Li--$^{87}$Rb system}
%\subsection{Motivation}

The Li--Rb system is extremely important both from the standpoint of ultracold atomic gases
and of ultracold molecular gases.  Not only are
$^6$Li and $^{87}$Rb in widespread use for studies of ultracold fermionic and bosonic
atomic gases, but LiRb has a relatively large dipole moment. It is therefore an important candidate for the study of ultracold
polar molecules and for the experimental study of electric-field-induced Feshbach resonances \cite{Li07}.  Therefore, understanding the low temperature collisional properties of this mixture is of great importance.

In 2005, Silber \emph{et al}.\ created a quantum degenerate Bose-Fermi mixture of 
$^6$Li and $^{87}$Rb
atoms in a magnetic trap with rubidium serving as the refrigerant \cite{Silber05}.
This experiment revealed the challenges of this approach to cooling lithium due to small magnitude of the interspecies scattering length at low magnetic fields.  Subsequently, inter-species Feshbach resonances in this system were found \cite{Deh08}.  Feshbach resonances may provide a means to enhance cooling in this mixture.  Inter-species resonances also provide a way to tune the interactions
in this Bose-Fermi mixture and may allow for the study of 
boson-mediated BCS pairing \cite{Bijlsma00}.
In addition, Feshbach resonances
may offer an efficient way of forming loosely bound LiRb dimers, which can then be
transferred from the excited vibrational state near threshold to the
ground vibrational state \cite{Sage05}.
In deeply bound vibrational states, the LiRb dimer 
has a large electric dipole moment (of up to 4.2 Debye) \cite{Aymar05}, 
and an ensemble of these molecules,
polarized by an external electric field, 
will interact strongly via the dipole-dipole interaction
which is both long range and anisotropic.
Such dipolar systems are predicted
to exhibit a wide variety of novel phenomena \cite{Santos00}
including superfluid, supersolid, Mott insulator, checkerboard, striped, and collapse phases for dipolar Bosonic gases \cite{Goral02,Yi07} as well as
novel superfluid phases of dipolar Fermi gases \cite{Damski03,Baranov04} and Luttinger liquid behavior in one dimensional traps \cite{Pedri08}.

Measurements of cross-thermalization in magnetically trapped $^6$Li--$^{87}$Rb and $^7$Li--$^{87}$Rb mixtures indicate that the interspecies triplet scattering lengths are $|a_{\mathrm{triplet}}^{6,87}| = 20^{+9}_{-6} \; \aB$ \cite{Silber05} and
$|a_{\mathrm{triplet}}^{7,87}| = 59^{+19}_{-19} \; \aB$ \cite{Marzok07}.  In addition, two heteronuclear Feshbach resonances were recently observed \cite{Deh08}.  The signs of the triplet scattering lengths and the location of Feshbach resonances in other atomic states, however, remains to be determined.  The sign of the scattering length is particularly important since it determines the global stability of this mixture.  Combining the experimental results we produce a new set of accurate LiRb interaction potentials which fully characterize the Li--Rb scattering properties in any combination of spin states and indicate that the sign of the $^6$Li--$^{87}$Rb triplet scattering length must be negative.  Using these potentials, we also predict the location and widths of all the Feshbach resonances below 2~kG for all Li--Rb spin combinations where $^{87}$Rb is in the lower hyperfine manifold.

\subsection{Interatomic Potentials}

Our starting point for this work is to model the triplet a$^3\Sigma$ and singlet X$^1\Sigma$ interaction potentials of LiRb by an analytical function of the form originally proposed by Degli-Esposti and Werner \cite{esposti:3351}
\be
V(R) = G(R) e^{-\alpha(R-R_c)} - T(R) \sum_{i=3}^{5} \frac{C_{2i}}{R^{2i}}
\ee
with
$G(R) = \sum_{l=0}^{8} g_l R^l$ and 
$T(R) = \frac{1}{2}(1+\tanh(1+t \,R))$.
The potential parameters were determined by varying this function to reproduce the overall shape and approximate number of bound states for the LiRb dimer expected from the \textit{ab initio} calculations \cite{Aymar05}.  The long range behavior is adjusted to match the van der Waals coefficient $C_6 = 2545 \,E_h \,\aB^6$ (where $E_h = 4.35974 \times 10^{-18}$~J) determined by Derevianko {\it et al}.\ \cite{Derevianko01}.

The amplitude and sign of the pure triplet and singlet $s$-wave scattering lengths as well as the positions of the Feshbach resonances are almost completely determined by the location of the least bound states of the potentials \cite{Weiner99}.  Since the long range behavior of the potentials has been accurately determined, the potentials can only be refined by making small adjustments to the short range repulsive wall while keeping the long range behavior fixed.
This fine tuning is done to simultaneously reproduce the experimentally measured 
triplet scattering length and Feshbach resonance locations and widths.

Since the full coupled channel calculation is computationally intensive, iteratively finding the proper modification of the model potentials to reproduce the experimentally observed resonances can be a lengthy process.
To simplify this search and to gain insight into the scattering properties of the Li--Rb system, we have employed the asymptotic bound state model ($\abm$) \cite{Moerdijk95, Stan04,Wille08} to first determine the energies of the last bound states of the triplet and singlet potentials consistent with the experimentally observed location of the Li--Rb Feshbach resonances.  We then tune the potential curves to reproduce these bound state energies.  Final refinement of the potentials is done to reproduce the exact location and widths of the observed Feshbach resonances.

\subsection{Results}

\begin{figure}[t!]
	\centering
	\includegraphics[width=9cm]{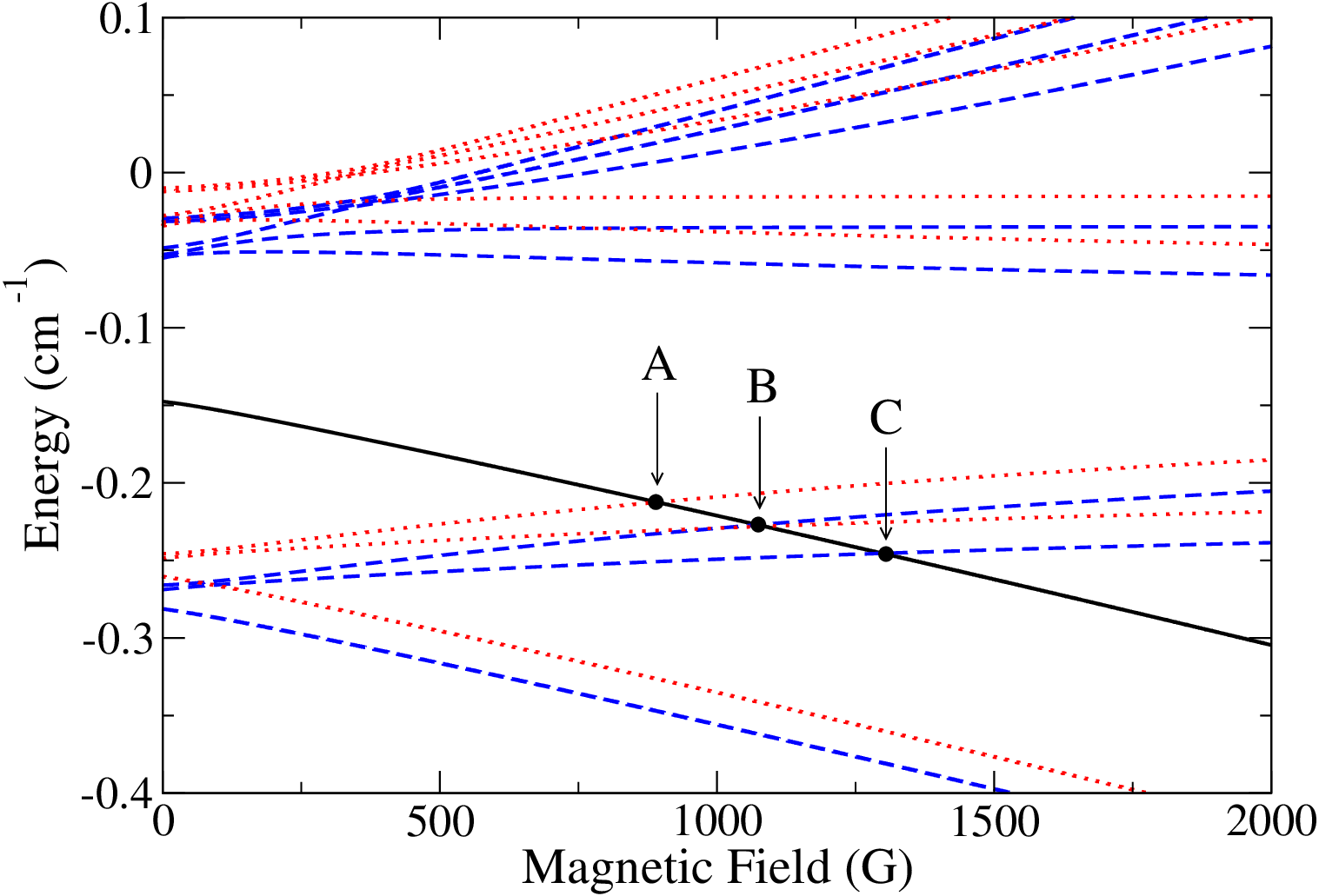}
	\renewcommand{\figurename}{Figure}
	\caption{Molecular bound state energies versus magnetic field computed with the asymptotic bound state ($\abm$) model. 
	The threshold for the $|\frac{1}{2},\frac{1}{2}\rangle_{^{6}\mathrm{Li}} \otimes |1,1\rangle_{^{87}\mathrm{Rb}}$
	  collision channel is shown by the solid line while the dashed (dotted) lines indicate the $s$-wave ($p$-wave) states.
	  These molecular state energies were computed given the least bound states, $E_S^{l}$, of the optimal singlet and triplet potentials of
	  $E_0^{l=0} = -0.106$~cm$^{-1}$, $E_0^{l=1} = -0.0870$~cm$^{-1}$, 
	  and
	  $E_1^{l=0} = -0.137$~cm$^{-1}$, $E_1^{l=1} = -0.116$~cm$^{-1}$.
	  The predicted resonance locations are close to the actual locations determined by the full coupled channel calculation 
	  and are indicated by the solid dots.  Near 890~G (A) the threshold crosses a $p$-wave molecular state 
	  and the corresponding $p$-wave elastic scattering cross section shown in Fig.\ \ref{fig:CCresults} is observed to rapidly diverge and then return to the background level. Likewise near 1070~G (B) the threshold crosses both a $p$-wave and $s$-wave molecular state and both the $s$-wave and $p$-wave elastic scattering cross sections are affected.  Finally, near 1300~G (C) a second $s$-wave induced Feshbach resonance occurs.}
	\label{fig:ABMLevels}
\end{figure}

The $\abm$ model is utilized and described elsewhere \cite{Wille08}, and here we provide a brief summary of the model and the details of our use of it to facilitate the search for the correct interaction potentials.  Feshbach resonances occur when a bound molecular state crosses the energy of the colliding atoms (open channel) at the dissociation threshold.  The $\abm$ model is therefore used to compute the energies of the molecular states (closest to the threshold) as a function of the magnetic field and to locate the crossings which result in Feshbach resonances.  Since the Hamiltonian (in the absence of magnetic dipole-dipole coupling) preserves the total spin angular momentum, 
only states with the same $M_F = M_S + M_{I_a} + M_{I_b}$ values as the initial state are considered.
The major simplifying assumption in the $\abm$ model is that the coupling between the channels (provided by the hyperfine interaction $\Vhf$) is small enough that the two-body bound states $\Psil$ can be represented to first order by \emph{uncoupled} orbital and spin states of the form $\Psil = \psils \otimes \alphastate$ where $\psils$ is the last bound state of either the pure singlet ($S=0$) or triplet ($S=1$) potential $V_S(R)$.  We therefore replace the spatial part of the 
Hamiltonian
(\ref{eq:Hrel}) with $\Hrel = \sum_{S,l}P_{S} E_S^{l}$ where $E_S^{l}$ are the energies of these last bound states.  In addition, we assume that the overlap of the singlet and triplet wave functions of the same orbital angular momentum $l$ is $\langle \psi_0^{l} | \psi_1^{l}\rangle = 1$, and the coupled bound state energies are found by diagonalizing the simplified Hamiltonian \cite{Wille08}.

Since the longrange part of the potential is known, the energies $E_S^{(l)}$ of the higher $l>0$ states are uniquely determined by the $l=0$ triplet $\ET = E_1^0$ and singlet $\ES = E_0^0$ energies.  $\ET$ and $\ES$ are therefore the only two free parameters in this model and they are adjusted until the threshold channel crosses the energy of the molecular states at the positions corresponding to the locations of the experimentally measured Feshbach resonances. Figure \ref{fig:ABMLevels} shows both the $s$ and $p$-wave molecular bound state energies versus magnetic field for all states with $M_F = 3/2$ computed within the asymptotic bound state model.  Although the $\abm$ model cannot predict the exact location of the Feshbach resonances, it does predict reliably the energies of the molecular channels in regions far from the crossings.  Therefore in the limit that the effect of the inter-state couplings on the energy is negligibly small, it provides an excellent estimate of the position of the Feshbach resonances.

\begin{figure}[t!]
	\centering
	\includegraphics[width=9cm]{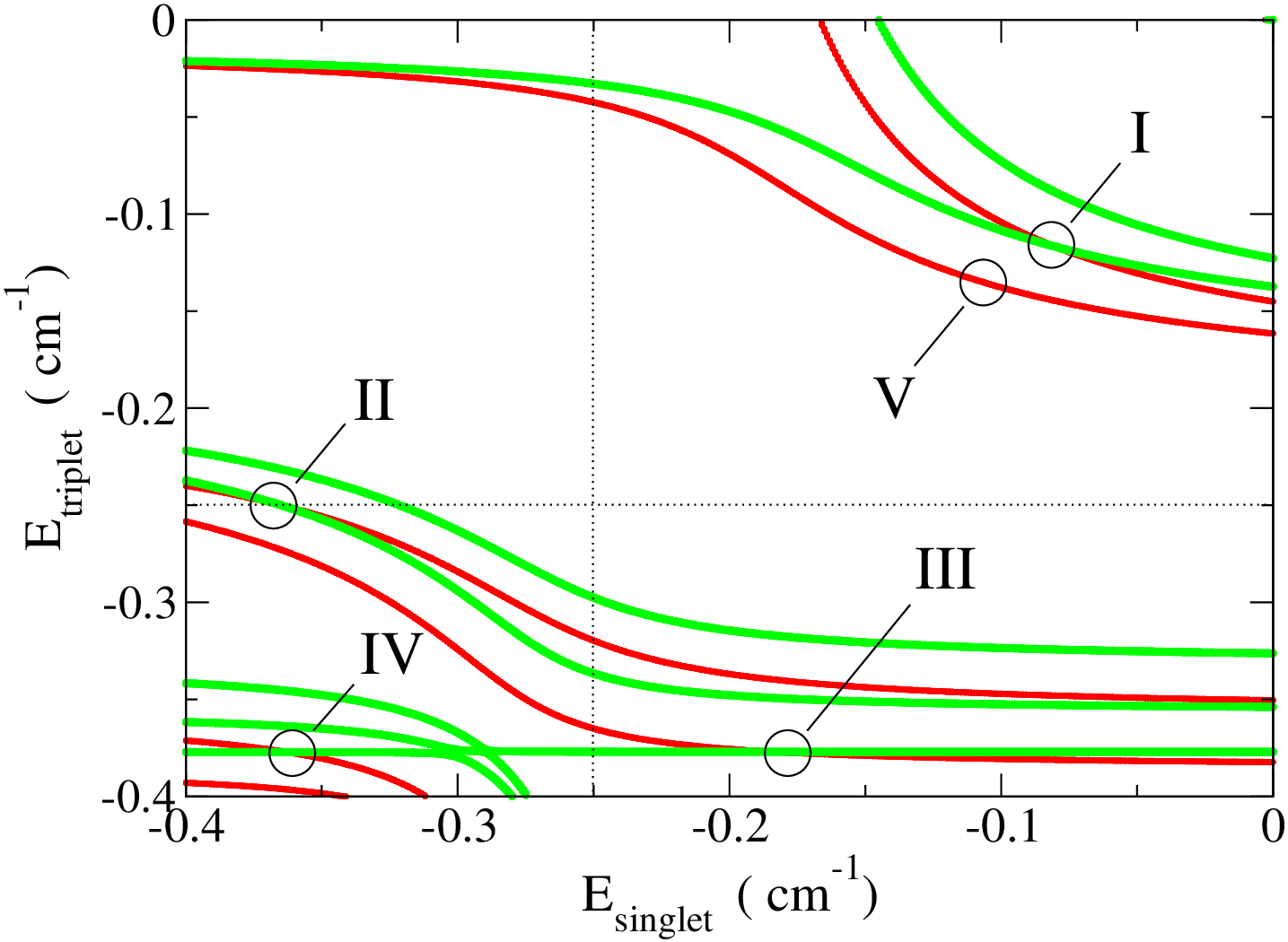}
	\caption{The locus of points in the $(\ES,\ET)$ parameter space where an 
	$s$-wave resonance occurs at one of the two experimentally determined locations 882.02~G (grey/green) or 1066.92~G (dark/red)
	for atoms in the 
         $|\frac{1}{2},\frac{1}{2}\rangle_{^{6}\mathrm{Li}} \otimes |1,1\rangle_{^{87}\mathrm{Rb}}$
         state.  The dotted lines indicate the approximate values for $\ES$ and $\ET$ beyond which a new bound state
         enters the potential at zero energy.  There are four regions (I-IV) indicated on the plot
         where an $s$-wave resonance occurs simultaneously at 882.02 and at 1066.92~G.  Region V
         indicates a range of values for which an $s$-wave resonance occurs at 1066.92 while a $p$-wave resonance
         (not represented in this plot) occurs at 882~G.  For each of these five candidate regions, the character of the predicted elastic cross
         	sections as a function of magnetic field was studied and the results of this analysis are discussed in the text.
	 }
	\label{fig:ResonancesVsEsEt}
\end{figure}

In order to utilize this model, we first employ it to determine the locus of points in the $(\ES,\ET)$ parameter space where an 
$s$-wave resonance occurs at one of the two experimentally determined locations 882.02~G or 1066.92~G
for atoms in the $|\frac{1}{2},\frac{1}{2}\rangle_{^{6}\mathrm{Li}} \otimes |1,1\rangle_{^{87}\mathrm{Rb}}$
state.  These points are plotted in Fig.\ \ref{fig:ResonancesVsEsEt}, and four regions (I-IV) indicated on the plot are found for which an $s$-wave resonance occurs simultaneously at 
882.02 and at 1066.92~G. Region V indicates a range of values $(\ES,\ET)$ for which an $s$-wave resonance occurs
at 1066.92~G and a $p$-wave resonance occurs at 882~G.  For each of the five candidate regions, the corresponding potentials
were generated and the predicted elastic scattering cross sections as a function of magnetic field were computed using the full coupled channel calculation.  In addition, the corresponding triplet scattering lengths were also computed.  Each of the four purely $s$-wave cases were ruled out based on a variety of reasons.  In region I, the lower resonance at 882~G is predicted to be a factor of 10 larger in width than the upper resonance at 1067~G in violation of the experimentally measured widths of 1.27~G and 10.62~G respectively \cite{Deh08}.  In region II, the relative widths of the resonances are (as in region I) incorrect and at these values for $(\ES,\ET)$ there would have been three additional and wide ($>$ 5~G) $s$-wave resonances below 200~G which were not observed in the experiment.  While in regions III and IV the ordering of the resonances is consistent with the experimental measurements (the upper resonance is larger than the lower resonance), in region III there is an additional wide ($>$10G) $s$-wave resonance below 200~G not observed in the experiment, and in region IV there is an additional $s$-wave resonance at approximately 960~G ($>$1G) in between the two observed resonances.  In addition, the triplet scattering length for regions III and IV corresponding to $\ET = -0.377$~cm$^{-1}$ is $a_{\mathrm{triplet}}^{6,87} = 105 \; \aB$.  This value is in disagreement with the experimentally determined value from measurements of the cross-thermalization in magnetically trapped $^6$Li--$^{87}$Rb mixtures which indicate that the interspecies triplet scattering length is $|a_{\mathrm{triplet}}^{6,87}| = 20^{+9}_{-6} \; \aB$ \cite{Silber05}.

In order to verify the robustness of these findings, we constructed a pair of Lennard-Jones potentials ($V(R) = C_{12}/R^{12} - C_{6}/R^{6}$) with the same $C_6$ coefficient, roughly the same number of bound states, and the same least bound state energies $(\ES,\ET)$ as the fitted potential in each of the regions considered.  Using these potentials, we verified that the Feshbach resonance locations and scattering lengths are essentially the same as for the fitted potentials and are insensitive to the short range details of the potentials.  This check provides an important verification of our characterization of the four purely $s$-wave candidate regions.  The conclusion is that the experimentally observed Feshbach resonances are inconsistent with pure $s$-wave resonances, and we must consider the possibility that at least one of the resonances originates from a $p$-wave molecular state.

\begin{figure}[ht]
  \begin{center}
    \includegraphics[width=0.5\textwidth,angle=0]{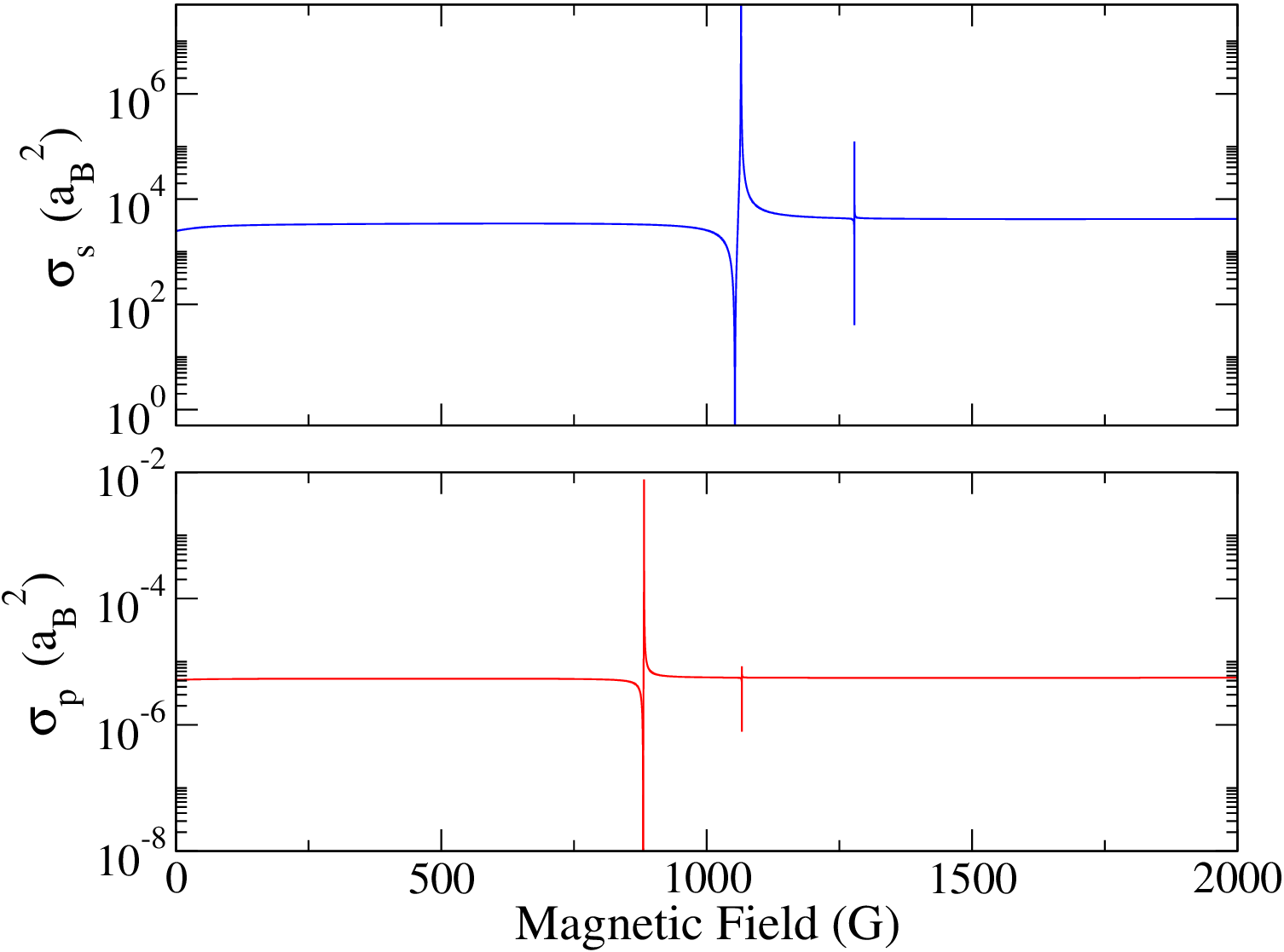}
     \caption{Magnetic field dependence of the $s$-wave (upper panel) and $p$-wave (lower panel) elastic scattering cross sections for atoms in the $|\frac{1}{2},\frac{1}{2}\rangle_{^{6}\mathrm{Li}} \otimes |1,1\rangle_{^{87}\mathrm{Rb}}$ state.  These results are from the coupled channel calculations for a collision energy of 144~nK and using the optimal singlet and triplet potentials.  Only the $m_l = 0$ contribution to the $p$-wave elastic scattering cross section is shown. Two $s$-wave resonances occur at 1065~G and 1278~G while two $p$-wave resonances
occur at 882~G and 1066~G.}
  \label{fig:CCresults}
  \end{center}
\end{figure}

Region V in Fig.\ \ref{fig:ResonancesVsEsEt} represents the only location in the $(\ES,\ET)$ parameter space for which only one $s$-wave resonance occurs below 1.2~kG (at 1067~G) and a $p$-wave resonance occurs at 882~G.  All other branches displayed in Fig.\ \ref{fig:ResonancesVsEsEt} involve at least one additional $s$-wave resonance occurring in a location where none was observed experimentally.  Along the locus of $(\ES,\ET)$ values for which these two resonances occur at the correct locations, an additional $p$-wave resonance was found to occur somewhere between 1081 and 1024~G while the width of the $s$-wave resonance at 1065~G was found to vary from 5~G to 35~G.  At the precise $(\ES,\ET)$ values for which the second $p$-wave resonance was coincident with the $s$-wave resonance at 1065~G, the $s$-wave resonance width was $\DB=11.53$~G, consistent with the experimentally measured value, $\DBexp=10.62$~G, for the full width at half maximum for the trap loss feature.  For these optimal singlet and triplet potentials, the bound state energies are $\ES = -0.106$~cm$^{-1}$ and $\ET = -0.137$~cm$^{-1}$.

Figure \ref{fig:CCresults} shows the 
result of the full coupled channel calculation performed using the refined potentials.
The elastic scattering cross-sections for the 
$|\frac{1}{2},\frac{1}{2}\rangle_{^{6}\mathrm{Li}} \otimes |1,1\rangle_{^{87}\mathrm{Rb}}$ 
state shows divergences at magnetic field values at 1065~G and 882~G, in excellent agreement
with the experimentally determined Feshbach resonance positions.  In addition, the triplet scattering length from the fine tuned triplet potential was found to be $a_{\mathrm{triplet}}^{6,87} = -19.8 \; \aB$, also in excellent agreement with the experimentally determined value.  
For the reduced mass corresponding to a 
$^7$Li--$^{87}$Rb complex, the optimal fine tuned triplet potential predicts
$a_{\mathrm{triplet}}^{7,87} = 448 \; \aB$, in disagreement with the experimental measurement of $|a_{\mathrm{triplet}}^{7,87}| = 59^{+19}_{-19} \; \aB$ \cite{Marzok07}.   Close inspection of this potential reveals that there is a bound state very close to the dissociation threshold for the $^7$Li--$^{87}$Rb triplet state.  In the case, a small uncertainty in the exact location of this very weakly bound state (arising from uncertainties in the exact shape of the potential) translates into a very large uncertainty in the predicted triplet scattering length for $^7$Li--$^{87}$Rb mixtures.  We verified the robustness of these results by changing the short range part of the potentials so that the number of bound states was different from the optimal potentials by more than 20\% while still producing the same energy of the least bound states.  As a result, the Feshbach resonance locations and scattering lengths did not change significantly.
In addition, we generated a set of Lennard-Jones potentials which reproduced the same least bound state energies and Feshbach resonance structure as the optimal fitted potentials. These potentials have a very different short range shape than the fitted potentials and they resulted in triplet scattering lengths of
$a_{\mathrm{triplet}}^{6,87} = -22.6 \; \aB$ and
$a_{\mathrm{triplet}}^{7,87} = -333 \; \aB$ confirming that the determination for the $^6$Li--$^{87}$Rb triplet scattering length is very reliable (independent of the details of the short range part of the potential) while the $^7$Li--$^{87}$Rb triplet scattering length cannot be reliably predicted given the proximity of a zero-energy resonance for this combination \cite{Arndt97}.  We note that if the triplet scattering length for the $^7$Li--$^{87}$Rb mixture is, in fact, negative, the experimental determination of its absolute magnitude can be complicated by the Ramsauer-Townsend effect in which the scattering cross section varies strongly with and may actually vanish at an experimentally relevant collision energy \cite{Aubin06}.

\begin{table}
\caption{Position and width of  $^6$Li--$^{87}$Rb Feshbach resonances for magnetic fields below 2~kG
determined from the coupled channel calculations.  The experimentally measured Feshbach resonances (and absence of resonances below 1.2~kG) are also included for comparison. The experimentally determined width $\DBexp$ is the full width at half maximum of the trap loss feature and, although related, it is not equivalent to $\DB$ (only defined for $s$-wave resonances).  Several resonances were found which exhibited a suppressed oscillation due to comparable coupling to inelastic channels \cite{Hutson07}
and could not be assigned a width in the usual way (in accordance with Eq.\ \ref{eq:ResForm}).  In these cases, the maximum and minimum elastic scattering lengths of the oscillation were
identified and the distance between them is indicated in parenthesis.}
\begin{ruledtabular}
\begin{tabular}{c c c c c c}
 \multicolumn{2}{c}{atomic states} &  \multicolumn{2}{c}{Theory} &  \multicolumn{2}{c}{Experiment \cite{Deh08}} \\
$|f,m_f\rangle_{6}$ & $|f,m_f\rangle_{87}$ & $\br$ & $\DB$ & $\br$ & $\DBexp$ \\
&   & (G) & (G) & (G) & (G) \\
  \hline
  \hline

  & & $882$ & $p$-wave & 882.02 & 1.27  \\
   &  & $1065$ & $11.5$ & 1066.92 & 10.62 \\
    \raisebox{1.5ex}[0pt]{$|\frac{1}{2},\frac{1}{2}\rangle$} &  \raisebox{1.5ex}[0pt]{$|1,1\rangle$} & $1066$ & $p$-wave & & \\
   &  & $1278$ & $0.07$ & & \\

  \hline
   & & $889$ & $p$-wave & & \\
   & & $1064$ & $17$ & & \\
   $|\frac{1}{2},\frac{1}{2}\rangle$ & $|1,0\rangle$ &  $1096$ & $p$-wave & \multicolumn{2}{c}{none below 1.2~kG}    \\
   &  & $1308.5$ & $(3)$ & &  \\
   &  & $1361.7$ & $p$-wave & & \\

  \hline

  \rule[-2mm]{0mm}{6mm} $|\frac{1}{2},\frac{1}{2}\rangle$ & $|1,-1\rangle$ & $1348$ & $(4)$ & \multicolumn{2}{c}{none below 1.2~kG}  \\

  \hline
  & & $773$ & $p$-wave & & \\
   &  & $923$ & $< 0.001$ & & \\
   &  & $926$ & $p$-wave & & \\
  \raisebox{1.5ex}[0pt]{$|\frac{1}{2},-\frac{1}{2}\rangle$} &  \raisebox{1.5ex}[0pt]{$|1,1\rangle$} & $1108.6$ & $11$ & & \\
   &  & $1119.5$ & $p$-wave & & \\
   &  & $1331$ & $0.08$ & & \\

% \raisebox{1.5ex}[0pt]{$|\frac{1}{2},\frac{1}{2}\rangle$} &  \raisebox{1.5ex}[0pt]{$|1,0\rangle$} 
 
  \hline
   & & $923$ & $p$-wave & & \\
   &  & $1105$ & $16.3$ & & \\
    \raisebox{1.5ex}[0pt]{$|\frac{1}{2},-\frac{1}{2}\rangle$} &  \raisebox{1.5ex}[0pt]{$|1,0\rangle$}  & $1150$ & $p$-wave & & \\
   &  & $1362$ & $(3)$ & & \\

  \hline
   & & $1408$ & $(4)$ & &  \\
    \raisebox{1.5ex}[0pt]{$|\frac{1}{2},-\frac{1}{2}\rangle$} &  \raisebox{1.5ex}[0pt]{$|1,-1\rangle$}   & $1611$ & $0.06$ & &\\

  \hline
  \rule[-2mm]{0mm}{6mm} $|\frac{3}{2},\frac{3}{2}\rangle$ & $|1,1\rangle$ & none &  &  \multicolumn{2}{c}{none below 1.2~kG}  \\

  \hline
  \rule[-2mm]{0mm}{6mm} $|\frac{3}{2},\frac{3}{2}\rangle$ & $|1,0\rangle$ & none & & &     \\

  \hline
  & & $953$ & $48.5$ & &  \\
    \raisebox{1.5ex}[0pt]{$|\frac{3}{2},\frac{3}{2}\rangle$} &  \raisebox{1.5ex}[0pt]{$|1,-1\rangle$}  & $1236.6$ & $p$-wave & & \\

  \hline
  & & $809$ & $p$-wave & &   \\
   &  & $960$ & $< 0.001$ & &  \\
     \raisebox{1.5ex}[0pt]{$|\frac{3}{2},-\frac{3}{2}\rangle$} &  \raisebox{1.5ex}[0pt]{$|1,1\rangle$}  & $971$ & $p$-wave & & \\
   &  & $1156$ & $11.7$ & & \\

  \hline
  & & $973$ & $p$-wave & &   \\
    \raisebox{1.5ex}[0pt]{$|\frac{3}{2},-\frac{3}{2}\rangle$} &  \raisebox{1.5ex}[0pt]{$|1,0\rangle$}  & $1149$ & $16.7$ & & \\

  \hline
  \rule[-2mm]{0mm}{6mm} $|\frac{3}{2},-\frac{3}{2}\rangle$ & $|1,-1\rangle$ & $1609$ & $0.07$  & & \\
\end{tabular}
\end{ruledtabular}
\label{tab:resonances}
\end{table}

Using the refined potentials, the $s$ and $p$ wave scattering cross sections as a function of magnetic field were calculated for all spin combinations where $^{87}$Rb is in the lower hyperfine manifold, and the location and widths of all resonances below 2~kG is summarized in Table \ref{tab:resonances}.
In experiments with $^{6}$Li--$^{87}$Rb mixtures, no Feshbach resonances were observed below 1.2~kG for the 
$|\frac{1}{2},\frac{1}{2}\rangle_{^{6}\mathrm{Li}} \otimes |1,0\rangle_{^{87}\mathrm{Rb}}$, 
$|\frac{1}{2},\frac{1}{2}\rangle_{^{6}\mathrm{Li}} \otimes |1,-1\rangle_{^{87}\mathrm{Rb}}$,
and
$|\frac{3}{2},\frac{3}{2}\rangle_{^{6}\mathrm{Li}} \otimes |1,1\rangle_{^{87}\mathrm{Rb}}$ states.  
The results presented in Table \ref{tab:resonances} are in agreement with the the last two of these observations but not the first.
It is possible that because the resonances present in the $|\frac{1}{2},\frac{1}{2}\rangle_{^{6}\mathrm{Li}} \otimes |1,0\rangle_{^{87}\mathrm{Rb}}$ combination are very similar in location and width to those of the $|\frac{1}{2},\frac{1}{2}\rangle_{^{6}\mathrm{Li}} \otimes |1,1\rangle_{^{87}\mathrm{Rb}}$ state that they may have been observed and erroneously concluded to arise from an impure state preparation.

\section{Conclusions}
We have presented calculations of experimentally relevant magnetic Feshbach resonances for both 
$^{85}$Rb--$^{87}$Rb and $^{6}$Li--$^{87}$Rb mixtures.  Our results build on recent experimental measurements of the triplet scattering lengths and Feshbach resonances in the Li--Rb system. We generate a set of refined LiRb interaction potentials which reproduce the location and widths of the measured resonances with high precision and use these refined potentials to predict additional experimentally relevant resonances for the $^{6}$Li--$^{87}$Rb mixture.  These potentials indicate that the $^{6}$Li--$^{87}$Rb triplet scattering length is $a_{\mathrm{triplet}}^{6,87} = -19.8 \; \aB$, consistent with cross-thermalization measurements.  We have verified that the predictions of these fine tuned potentials are robust in the sense that they only depend on the well-known longrange $C_6$ coefficient and are independent of both the details of the short range shape and the exact number of bound states of the interaction potentials.
Finally, we have identified a zero-energy resonance generated by a bound or virtual state very near to the dissociation threshold in the $^{7}$Li--$^{87}$Rb triplet state.  In the case of $^{85}$Rb--$^{87}$Rb collisions, using previously validated Rb$_2$ interaction potentials, we have conducted a search for low field Feshbach resonances which would be candidates for the realization of RF induced Feshbach resonances \cite{Hofferberth06}.  The significance of this work is that low field resonances, below 10~G, do exist for this mixture. These particular resonances are therefore very accessible and useful for chip based magnetic trap experiments.  In addition, they may be eventually very useful for the generation of spin state entanglement in optical lattices or atom chip magnetic traps \cite{Treutlein06}.

\begin{acknowledgments}
We thank Roman Krems for simulating and truly essential discussions and for his careful reading of this manuscript.  We also acknowledge Eite Tiesinga for providing the Rb$_2$ potentials and for his insights and advice on the use and utility of the $\abm$ model.  This work was supported by the Natural Sciences and Engineering Research Council of Canada (NSERC), the Canadian Foundation for Innovation (CFI), and the Canadian Institute for Advanced Research (CIfAR).
\end{acknowledgments}

%\bibliography{FeshbachPaperBib}% Produces the bibliography via BibTeX.

\begin{thebibliography}{45}
\expandafter\ifx\csname natexlab\endcsname\relax\def\natexlab#1{#1}\fi
\expandafter\ifx\csname bibnamefont\endcsname\relax
  \def\bibnamefont#1{#1}\fi
\expandafter\ifx\csname bibfnamefont\endcsname\relax
  \def\bibfnamefont#1{#1}\fi
\expandafter\ifx\csname citenamefont\endcsname\relax
  \def\citenamefont#1{#1}\fi
\expandafter\ifx\csname url\endcsname\relax
  \def\url#1{\texttt{#1}}\fi
\expandafter\ifx\csname urlprefix\endcsname\relax\def\urlprefix{URL }\fi
\providecommand{\bibinfo}[2]{#2}
\providecommand{\eprint}[2][]{\url{#2}}

\bibitem[{\citenamefont{Tiesinga et~al.}(1993)\citenamefont{Tiesinga, Verhaar,
  and Stoof}}]{Tiesinga93}
\bibinfo{author}{\bibfnamefont{E.}~\bibnamefont{Tiesinga}},
  \bibinfo{author}{\bibfnamefont{B.}~\bibnamefont{Verhaar}}, \bibnamefont{and}
  \bibinfo{author}{\bibfnamefont{H.}~\bibnamefont{Stoof}},
  \bibinfo{journal}{Phys. Rev. A} \textbf{\bibinfo{volume}{47}},
  \bibinfo{pages}{4114} (\bibinfo{year}{1993}).

\bibitem[{\citenamefont{Weiner et~al.}(1999)\citenamefont{Weiner, Bagnato,
  Zillo, and Julienne}}]{Weiner99}
\bibinfo{author}{\bibfnamefont{J.}~\bibnamefont{Weiner}},
  \bibinfo{author}{\bibfnamefont{V.}~\bibnamefont{Bagnato}},
  \bibinfo{author}{\bibfnamefont{S.}~\bibnamefont{Zillo}}, \bibnamefont{and}
  \bibinfo{author}{\bibfnamefont{P.}~\bibnamefont{Julienne}},
  \bibinfo{journal}{Rev. Mod. Phys.} \textbf{\bibinfo{volume}{71}},
  \bibinfo{pages}{1} (\bibinfo{year}{1999}).

\bibitem[{\citenamefont{Jones et~al.}(2006)\citenamefont{Jones, Tiesinga, Lett,
  and Julienne}}]{Jones06}
\bibinfo{author}{\bibfnamefont{K.~M.} \bibnamefont{Jones}},
  \bibinfo{author}{\bibfnamefont{E.}~\bibnamefont{Tiesinga}},
  \bibinfo{author}{\bibfnamefont{P.~D.} \bibnamefont{Lett}}, \bibnamefont{and}
  \bibinfo{author}{\bibfnamefont{P.~S.} \bibnamefont{Julienne}},
  \bibinfo{journal}{Rev. Mod. Phys.} \textbf{\bibinfo{volume}{78}},
  \bibinfo{pages}{483} (\bibinfo{year}{2006}).

\bibitem[{\citenamefont{Inouye et~al.}(1998)\citenamefont{Inouye, Andrews,
  Stenger, Miesner, Stamper-Kurn, and Ketterle}}]{Inouye98}
\bibinfo{author}{\bibfnamefont{S.}~\bibnamefont{Inouye}},
  \bibinfo{author}{\bibfnamefont{M.}~\bibnamefont{Andrews}},
  \bibinfo{author}{\bibfnamefont{J.}~\bibnamefont{Stenger}},
  \bibinfo{author}{\bibfnamefont{H.-J.} \bibnamefont{Miesner}},
  \bibinfo{author}{\bibfnamefont{D.}~\bibnamefont{Stamper-Kurn}},
  \bibnamefont{and} \bibinfo{author}{\bibfnamefont{W.}~\bibnamefont{Ketterle}},
  \bibinfo{journal}{Nature} \textbf{\bibinfo{volume}{392}},
  \bibinfo{pages}{151} (\bibinfo{year}{1998}).

\bibitem[{\citenamefont{Donley et~al.}(2002)\citenamefont{Donley, Claussen,
  Thompson, and Wieman}}]{Donley02}
\bibinfo{author}{\bibfnamefont{E.}~\bibnamefont{Donley}},
  \bibinfo{author}{\bibfnamefont{N.}~\bibnamefont{Claussen}},
  \bibinfo{author}{\bibfnamefont{S.}~\bibnamefont{Thompson}}, \bibnamefont{and}
  \bibinfo{author}{\bibfnamefont{C.}~\bibnamefont{Wieman}},
  \bibinfo{journal}{Nature} \textbf{\bibinfo{volume}{417}},
  \bibinfo{pages}{529} (\bibinfo{year}{2002}).

\bibitem[{\citenamefont{Claussen et~al.}(2002)\citenamefont{Claussen, Donley,
  Thompson, and Wieman}}]{Claussen02}
\bibinfo{author}{\bibfnamefont{N.}~\bibnamefont{Claussen}},
  \bibinfo{author}{\bibfnamefont{E.}~\bibnamefont{Donley}},
  \bibinfo{author}{\bibfnamefont{S.}~\bibnamefont{Thompson}}, \bibnamefont{and}
  \bibinfo{author}{\bibfnamefont{C.}~\bibnamefont{Wieman}},
  \bibinfo{journal}{Phys. Rev. Lett.} \textbf{\bibinfo{volume}{89}},
  \bibinfo{pages}{010401} (\bibinfo{year}{2002}).

\bibitem[{\citenamefont{Strecker et~al.}(2002)\citenamefont{Strecker,
  Partridge, Truscott, and Hulet}}]{Strecker02}
\bibinfo{author}{\bibfnamefont{K.}~\bibnamefont{Strecker}},
  \bibinfo{author}{\bibfnamefont{G.}~\bibnamefont{Partridge}},
  \bibinfo{author}{\bibfnamefont{A.}~\bibnamefont{Truscott}}, \bibnamefont{and}
  \bibinfo{author}{\bibfnamefont{R.}~\bibnamefont{Hulet}},
  \bibinfo{journal}{Nature} \textbf{\bibinfo{volume}{417}},
  \bibinfo{pages}{150} (\bibinfo{year}{2002}).

\bibitem[{\citenamefont{Khaykovich et~al.}(2002)\citenamefont{Khaykovich,
  Schreck, Ferrari, Bourdel, Cubizolles, Carr, Castin, and
  Salomon}}]{Khaykovich02}
\bibinfo{author}{\bibfnamefont{L.}~\bibnamefont{Khaykovich}},
  \bibinfo{author}{\bibfnamefont{F.}~\bibnamefont{Schreck}},
  \bibinfo{author}{\bibfnamefont{G.}~\bibnamefont{Ferrari}},
  \bibinfo{author}{\bibfnamefont{T.}~\bibnamefont{Bourdel}},
  \bibinfo{author}{\bibfnamefont{J.}~\bibnamefont{Cubizolles}},
  \bibinfo{author}{\bibfnamefont{L.~D.} \bibnamefont{Carr}},
  \bibinfo{author}{\bibfnamefont{Y.}~\bibnamefont{Castin}}, \bibnamefont{and}
  \bibinfo{author}{\bibfnamefont{C.}~\bibnamefont{Salomon}},
  \bibinfo{journal}{Science} \textbf{\bibinfo{volume}{296}},
  \bibinfo{pages}{1290} (\bibinfo{year}{2002}).

\bibitem[{\citenamefont{Chin et~al.}(2004)\citenamefont{Chin, Bartenstein,
  Altmeyer, Riedl, Jochim, Denschlag, and Grimm}}]{Chin04}
\bibinfo{author}{\bibfnamefont{C.}~\bibnamefont{Chin}},
  \bibinfo{author}{\bibfnamefont{M.}~\bibnamefont{Bartenstein}},
  \bibinfo{author}{\bibfnamefont{A.}~\bibnamefont{Altmeyer}},
  \bibinfo{author}{\bibfnamefont{S.}~\bibnamefont{Riedl}},
  \bibinfo{author}{\bibfnamefont{S.}~\bibnamefont{Jochim}},
  \bibinfo{author}{\bibfnamefont{J.~H.} \bibnamefont{Denschlag}},
  \bibnamefont{and} \bibinfo{author}{\bibfnamefont{R.}~\bibnamefont{Grimm}},
  \bibinfo{journal}{Science} \textbf{\bibinfo{volume}{305}},
  \bibinfo{pages}{1128} (\bibinfo{year}{2004}).

\bibitem[{\citenamefont{Zwierlein et~al.}(2004)\citenamefont{Zwierlein, Stan,
  Schunck, Raupach, Kerman, and Ketterle}}]{Zwierlein04}
\bibinfo{author}{\bibfnamefont{M.}~\bibnamefont{Zwierlein}},
  \bibinfo{author}{\bibfnamefont{C.}~\bibnamefont{Stan}},
  \bibinfo{author}{\bibfnamefont{C.}~\bibnamefont{Schunck}},
  \bibinfo{author}{\bibfnamefont{S.}~\bibnamefont{Raupach}},
  \bibinfo{author}{\bibfnamefont{A.}~\bibnamefont{Kerman}}, \bibnamefont{and}
  \bibinfo{author}{\bibfnamefont{W.}~\bibnamefont{Ketterle}},
  \bibinfo{journal}{Phys. Rev. Lett.} \textbf{\bibinfo{volume}{92}},
  \bibinfo{pages}{120403} (\bibinfo{year}{2004}).

\bibitem[{\citenamefont{Regal et~al.}(2004)\citenamefont{Regal, Greiner, and
  Jin}}]{Regal04}
\bibinfo{author}{\bibfnamefont{C.}~\bibnamefont{Regal}},
  \bibinfo{author}{\bibfnamefont{M.}~\bibnamefont{Greiner}}, \bibnamefont{and}
  \bibinfo{author}{\bibfnamefont{D.}~\bibnamefont{Jin}},
  \bibinfo{journal}{Phys. Rev. Lett.} \textbf{\bibinfo{volume}{92}},
  \bibinfo{pages}{040403} (\bibinfo{year}{2004}).

\bibitem[{\citenamefont{Bhattacharya et~al.}(2004)\citenamefont{Bhattacharya,
  Baksmaty, Weiss, and Bigelow}}]{Bhattacharya04}
\bibinfo{author}{\bibfnamefont{M.}~\bibnamefont{Bhattacharya}},
  \bibinfo{author}{\bibfnamefont{L.}~\bibnamefont{Baksmaty}},
  \bibinfo{author}{\bibfnamefont{S.}~\bibnamefont{Weiss}}, \bibnamefont{and}
  \bibinfo{author}{\bibfnamefont{N.}~\bibnamefont{Bigelow}},
  \bibinfo{journal}{Eur. Phys. J. D} \textbf{\bibinfo{volume}{31}},
  \bibinfo{pages}{301} (\bibinfo{year}{2004}).

\bibitem[{\citenamefont{Stan et~al.}(2004)\citenamefont{Stan, Zwierlein,
  Schunck, Raupach, and Ketterle}}]{Stan04}
\bibinfo{author}{\bibfnamefont{C.~A.} \bibnamefont{Stan}},
  \bibinfo{author}{\bibfnamefont{M.~W.} \bibnamefont{Zwierlein}},
  \bibinfo{author}{\bibfnamefont{C.~H.} \bibnamefont{Schunck}},
  \bibinfo{author}{\bibfnamefont{S.~M.~F.} \bibnamefont{Raupach}},
  \bibnamefont{and} \bibinfo{author}{\bibfnamefont{W.}~\bibnamefont{Ketterle}},
  \bibinfo{journal}{Phys. Rev. Lett.} \textbf{\bibinfo{volume}{93}},
  \bibinfo{pages}{143001} (\bibinfo{year}{2004}).

\bibitem[{\citenamefont{Inouye et~al.}(2004)\citenamefont{Inouye, Goldwin,
  Olsen, C.Ticknor, Bohn, and Jin}}]{Inouye04}
\bibinfo{author}{\bibfnamefont{S.}~\bibnamefont{Inouye}},
  \bibinfo{author}{\bibfnamefont{J.}~\bibnamefont{Goldwin}},
  \bibinfo{author}{\bibfnamefont{M.}~\bibnamefont{Olsen}},
  \bibinfo{author}{\bibnamefont{C.Ticknor}},
  \bibinfo{author}{\bibfnamefont{J.}~\bibnamefont{Bohn}}, \bibnamefont{and}
  \bibinfo{author}{\bibfnamefont{D.}~\bibnamefont{Jin}},
  \bibinfo{journal}{Phys. Rev. Lett.} \textbf{\bibinfo{volume}{93}},
  \bibinfo{pages}{183201} (\bibinfo{year}{2004}).

\bibitem[{\citenamefont{Papp and Wieman}(2006)}]{Papp06}
\bibinfo{author}{\bibfnamefont{S.~B.} \bibnamefont{Papp}} \bibnamefont{and}
  \bibinfo{author}{\bibfnamefont{C.~E.} \bibnamefont{Wieman}},
  \bibinfo{journal}{Phys. Rev. Lett.} \textbf{\bibinfo{volume}{97}},
  \bibinfo{pages}{180404} (\bibinfo{year}{2006}).

\bibitem[{\citenamefont{Deh et~al.}(2008)\citenamefont{Deh, Marzok, Zimmermann,
  and Courteille}}]{Deh08}
\bibinfo{author}{\bibfnamefont{B.}~\bibnamefont{Deh}},
  \bibinfo{author}{\bibfnamefont{C.}~\bibnamefont{Marzok}},
  \bibinfo{author}{\bibfnamefont{C.}~\bibnamefont{Zimmermann}},
  \bibnamefont{and}
  \bibinfo{author}{\bibfnamefont{P.}~\bibnamefont{Courteille}},
  \bibinfo{journal}{Phys. Rev. A} \textbf{\bibinfo{volume}{77}},
  \bibinfo{pages}{010701(R)} (\bibinfo{year}{2008}).

\bibitem[{\citenamefont{Wille et~al.}(2008)\citenamefont{Wille, Spiegelhalder,
  Kerner, Naik, Trenkwalder, Hendl, Schreck, Grimm, Tiecke, Walraven
  et~al.}}]{Wille08}
\bibinfo{author}{\bibfnamefont{E.}~\bibnamefont{Wille}},
  \bibinfo{author}{\bibfnamefont{F.~M.} \bibnamefont{Spiegelhalder}},
  \bibinfo{author}{\bibfnamefont{G.}~\bibnamefont{Kerner}},
  \bibinfo{author}{\bibfnamefont{D.}~\bibnamefont{Naik}},
  \bibinfo{author}{\bibfnamefont{A.}~\bibnamefont{Trenkwalder}},
  \bibinfo{author}{\bibfnamefont{G.}~\bibnamefont{Hendl}},
  \bibinfo{author}{\bibfnamefont{F.}~\bibnamefont{Schreck}},
  \bibinfo{author}{\bibfnamefont{R.}~\bibnamefont{Grimm}},
  \bibinfo{author}{\bibfnamefont{T.~G.} \bibnamefont{Tiecke}},
  \bibinfo{author}{\bibfnamefont{J.~T.~M.} \bibnamefont{Walraven}},
  \bibnamefont{et~al.}, \bibinfo{journal}{Phys. Rev. Lett.}
  \textbf{\bibinfo{volume}{100}}, \bibinfo{pages}{053201}
  (\bibinfo{year}{2008}).

\bibitem[{\citenamefont{Heiselberg et~al.}(2000)\citenamefont{Heiselberg,
  Pethick, Smith, and Viverit}}]{Heiselberg00}
\bibinfo{author}{\bibfnamefont{H.}~\bibnamefont{Heiselberg}},
  \bibinfo{author}{\bibfnamefont{C.}~\bibnamefont{Pethick}},
  \bibinfo{author}{\bibfnamefont{H.}~\bibnamefont{Smith}}, \bibnamefont{and}
  \bibinfo{author}{\bibfnamefont{L.}~\bibnamefont{Viverit}},
  \bibinfo{journal}{Phys. Rev. Lett.} \textbf{\bibinfo{volume}{85}},
  \bibinfo{pages}{2418} (\bibinfo{year}{2000}).

\bibitem[{\citenamefont{Buchler and Blatter}(2003)}]{Buchler03}
\bibinfo{author}{\bibfnamefont{H.}~\bibnamefont{Buchler}} \bibnamefont{and}
  \bibinfo{author}{\bibfnamefont{G.}~\bibnamefont{Blatter}},
  \bibinfo{journal}{Phys. Rev. Lett.} \textbf{\bibinfo{volume}{91}},
  \bibinfo{pages}{130404} (\bibinfo{year}{2003}).

\bibitem[{\citenamefont{Sage et~al.}(2005)\citenamefont{Sage, Sainis, Bergeman,
  and DeMille}}]{Sage05}
\bibinfo{author}{\bibfnamefont{J.}~\bibnamefont{Sage}},
  \bibinfo{author}{\bibfnamefont{S.}~\bibnamefont{Sainis}},
  \bibinfo{author}{\bibfnamefont{T.}~\bibnamefont{Bergeman}}, \bibnamefont{and}
  \bibinfo{author}{\bibfnamefont{D.}~\bibnamefont{DeMille}},
  \bibinfo{journal}{Phys. Rev. Lett.} \textbf{\bibinfo{volume}{94}},
  \bibinfo{pages}{203001} (\bibinfo{year}{2005}).

\bibitem[{\citenamefont{Tiesinga}(2008)}]{Tiesinga08}
\bibinfo{author}{\bibfnamefont{E.}~\bibnamefont{Tiesinga}}
  (\bibinfo{year}{2008}), \bibinfo{note}{private communication}.

\bibitem[{\citenamefont{Jr. et~al.}(1998)\citenamefont{Jr., Bohn, Esry, and
  Greene}}]{Burke98}
\bibinfo{author}{\bibfnamefont{J.~P.~B.} \bibnamefont{Jr.}},
  \bibinfo{author}{\bibfnamefont{J.~L.} \bibnamefont{Bohn}},
  \bibinfo{author}{\bibfnamefont{B.~D.} \bibnamefont{Esry}}, \bibnamefont{and}
  \bibinfo{author}{\bibfnamefont{C.~H.} \bibnamefont{Greene}},
  \bibinfo{journal}{Phys. Rev. Lett.} \textbf{\bibinfo{volume}{80}},
  \bibinfo{pages}{2097} (\bibinfo{year}{1998}).

\bibitem[{\citenamefont{M.Aymar and O.Dulieu}(2005)}]{Aymar05}
\bibinfo{author}{\bibnamefont{M.Aymar}} \bibnamefont{and}
  \bibinfo{author}{\bibnamefont{O.Dulieu}}, \bibinfo{journal}{J. Chem. Phys.}
  \textbf{\bibinfo{volume}{122}}, \bibinfo{pages}{204302}
  (\bibinfo{year}{2005}).

\bibitem[{\citenamefont{Derevianko et~al.}(2001)\citenamefont{Derevianko, Babb,
  and Dalgarno}}]{Derevianko01}
\bibinfo{author}{\bibfnamefont{A.}~\bibnamefont{Derevianko}},
  \bibinfo{author}{\bibfnamefont{J.~F.} \bibnamefont{Babb}}, \bibnamefont{and}
  \bibinfo{author}{\bibfnamefont{A.}~\bibnamefont{Dalgarno}},
  \bibinfo{journal}{Phys. Rev. A} \textbf{\bibinfo{volume}{63}},
  \bibinfo{pages}{052704} (\bibinfo{year}{2001}).

\bibitem[{\citenamefont{Moerdijk et~al.}(1995)\citenamefont{Moerdijk, Verhaar,
  and Axelsson}}]{Moerdijk95}
\bibinfo{author}{\bibfnamefont{A.~J.} \bibnamefont{Moerdijk}},
  \bibinfo{author}{\bibfnamefont{B.~J.} \bibnamefont{Verhaar}},
  \bibnamefont{and} \bibinfo{author}{\bibfnamefont{A.}~\bibnamefont{Axelsson}},
  \bibinfo{journal}{Phys. Rev. A} \textbf{\bibinfo{volume}{51}},
  \bibinfo{pages}{4852} (\bibinfo{year}{1995}).

\bibitem[{\citenamefont{Li and Krems}(2007)}]{Li07}
\bibinfo{author}{\bibfnamefont{Z.}~\bibnamefont{Li}} \bibnamefont{and}
  \bibinfo{author}{\bibfnamefont{R.}~\bibnamefont{Krems}},
  \bibinfo{journal}{Phys. Rev. A} \textbf{\bibinfo{volume}{75}},
  \bibinfo{pages}{032709} (\bibinfo{year}{2007}).

\bibitem[{\citenamefont{Krems and Dalgarno}(2004)}]{Krems04b}
\bibinfo{author}{\bibfnamefont{R.~V.} \bibnamefont{Krems}} \bibnamefont{and}
  \bibinfo{author}{\bibfnamefont{A.}~\bibnamefont{Dalgarno}},
  \bibinfo{journal}{J. Chem. Phys.} \textbf{\bibinfo{volume}{120}},
  \bibinfo{pages}{2296} (\bibinfo{year}{2004}).

\bibitem[{\citenamefont{Hutson}(2007)}]{Hutson07}
\bibinfo{author}{\bibfnamefont{J.~M.} \bibnamefont{Hutson}},
  \bibinfo{journal}{New J. Phys.} \textbf{\bibinfo{volume}{9}},
  \bibinfo{pages}{152} (\bibinfo{year}{2007}).

\bibitem[{\citenamefont{Krems}(2004)}]{Krems04a}
\bibinfo{author}{\bibfnamefont{R.~V.} \bibnamefont{Krems}},
  \bibinfo{journal}{Phys. Rev. Lett.} \textbf{\bibinfo{volume}{93}},
  \bibinfo{pages}{013201} (\bibinfo{year}{2004}).

\bibitem[{\citenamefont{Hofferberth et~al.}(2006)\citenamefont{Hofferberth,
  Lesanovsky, Fischer, Verdu, and Schmiedmayer}}]{Hofferberth06}
\bibinfo{author}{\bibfnamefont{S.}~\bibnamefont{Hofferberth}},
  \bibinfo{author}{\bibfnamefont{I.}~\bibnamefont{Lesanovsky}},
  \bibinfo{author}{\bibfnamefont{B.}~\bibnamefont{Fischer}},
  \bibinfo{author}{\bibfnamefont{J.}~\bibnamefont{Verdu}}, \bibnamefont{and}
  \bibinfo{author}{\bibfnamefont{J.}~\bibnamefont{Schmiedmayer}},
  \bibinfo{journal}{Nat. Phys.} \textbf{\bibinfo{volume}{2}},
  \bibinfo{pages}{710} (\bibinfo{year}{2006}).

\bibitem[{\citenamefont{Widera et~al.}(2004)\citenamefont{Widera, Mandel,
  Greiner, Kreim, H{\"a}nsch, and Bloch}}]{Widera04}
\bibinfo{author}{\bibfnamefont{A.}~\bibnamefont{Widera}},
  \bibinfo{author}{\bibfnamefont{O.}~\bibnamefont{Mandel}},
  \bibinfo{author}{\bibfnamefont{M.}~\bibnamefont{Greiner}},
  \bibinfo{author}{\bibfnamefont{S.}~\bibnamefont{Kreim}},
  \bibinfo{author}{\bibfnamefont{T.~W.} \bibnamefont{H{\"a}nsch}},
  \bibnamefont{and} \bibinfo{author}{\bibfnamefont{I.}~\bibnamefont{Bloch}},
  \bibinfo{journal}{Phys. Rev. Lett.} \textbf{\bibinfo{volume}{92}},
  \bibinfo{pages}{160406} (\bibinfo{year}{2004}).

\bibitem[{\citenamefont{Erhard et~al.}(2004)\citenamefont{Erhard, Schmaljohann,
  Kronj{\"a}ger, Bongs, and Sengstock}}]{Erhard04}
\bibinfo{author}{\bibfnamefont{M.}~\bibnamefont{Erhard}},
  \bibinfo{author}{\bibfnamefont{H.}~\bibnamefont{Schmaljohann}},
  \bibinfo{author}{\bibfnamefont{J.}~\bibnamefont{Kronj{\"a}ger}},
  \bibinfo{author}{\bibfnamefont{K.}~\bibnamefont{Bongs}}, \bibnamefont{and}
  \bibinfo{author}{\bibfnamefont{K.}~\bibnamefont{Sengstock}},
  \bibinfo{journal}{Phys. Rev. A} \textbf{\bibinfo{volume}{69}},
  \bibinfo{pages}{032705} (\bibinfo{year}{2004}).

\bibitem[{\citenamefont{Silber et~al.}(2005)\citenamefont{Silber, G{\"u}nther,
  Marzok, Deh, Courteille, and Zimmermann}}]{Silber05}
\bibinfo{author}{\bibfnamefont{C.}~\bibnamefont{Silber}},
  \bibinfo{author}{\bibfnamefont{S.}~\bibnamefont{G{\"u}nther}},
  \bibinfo{author}{\bibfnamefont{C.}~\bibnamefont{Marzok}},
  \bibinfo{author}{\bibfnamefont{B.}~\bibnamefont{Deh}},
  \bibinfo{author}{\bibfnamefont{P.}~\bibnamefont{Courteille}},
  \bibnamefont{and}
  \bibinfo{author}{\bibfnamefont{C.}~\bibnamefont{Zimmermann}},
  \bibinfo{journal}{Phys. Rev. Lett.} \textbf{\bibinfo{volume}{95}},
  \bibinfo{pages}{170408} (\bibinfo{year}{2005}).

\bibitem[{\citenamefont{Bijlsma et~al.}(2000)\citenamefont{Bijlsma, Heringa,
  and Stoof}}]{Bijlsma00}
\bibinfo{author}{\bibfnamefont{M.~J.} \bibnamefont{Bijlsma}},
  \bibinfo{author}{\bibfnamefont{B.~A.} \bibnamefont{Heringa}},
  \bibnamefont{and} \bibinfo{author}{\bibfnamefont{H.~T.~C.}
  \bibnamefont{Stoof}}, \bibinfo{journal}{Phys. Rev. A}
  \textbf{\bibinfo{volume}{61}}, \bibinfo{pages}{053601}
  (\bibinfo{year}{2000}).

\bibitem[{\citenamefont{Santos et~al.}(2000)\citenamefont{Santos, Shlyapnikov,
  Zoller, and Lewenstein}}]{Santos00}
\bibinfo{author}{\bibfnamefont{L.}~\bibnamefont{Santos}},
  \bibinfo{author}{\bibfnamefont{G.~V.} \bibnamefont{Shlyapnikov}},
  \bibinfo{author}{\bibfnamefont{P.}~\bibnamefont{Zoller}}, \bibnamefont{and}
  \bibinfo{author}{\bibfnamefont{M.}~\bibnamefont{Lewenstein}},
  \bibinfo{journal}{Phys. Rev. Lett.} \textbf{\bibinfo{volume}{85}},
  \bibinfo{pages}{1791} (\bibinfo{year}{2000}).

\bibitem[{\citenamefont{G\'oral et~al.}(2002)\citenamefont{G\'oral, Santos, and
  Lewenstein}}]{Goral02}
\bibinfo{author}{\bibfnamefont{K.}~\bibnamefont{G\'oral}},
  \bibinfo{author}{\bibfnamefont{L.}~\bibnamefont{Santos}}, \bibnamefont{and}
  \bibinfo{author}{\bibfnamefont{M.}~\bibnamefont{Lewenstein}},
  \bibinfo{journal}{Phys. Rev. Lett.} \textbf{\bibinfo{volume}{88}},
  \bibinfo{pages}{170406} (\bibinfo{year}{2002}).

\bibitem[{\citenamefont{Yi et~al.}(2007)\citenamefont{Yi, Li, and Sun}}]{Yi07}
\bibinfo{author}{\bibfnamefont{S.}~\bibnamefont{Yi}},
  \bibinfo{author}{\bibfnamefont{T.}~\bibnamefont{Li}}, \bibnamefont{and}
  \bibinfo{author}{\bibfnamefont{C.}~\bibnamefont{Sun}},
  \bibinfo{journal}{Phys. Rev. Lett.} \textbf{\bibinfo{volume}{98}},
  \bibinfo{pages}{260405} (\bibinfo{year}{2007}).

\bibitem[{\citenamefont{Damski et~al.}(2003)\citenamefont{Damski, Santos,
  Tiemann, Lewenstein, Kotochigova, Julienne, and Zoller}}]{Damski03}
\bibinfo{author}{\bibfnamefont{B.}~\bibnamefont{Damski}},
  \bibinfo{author}{\bibfnamefont{L.}~\bibnamefont{Santos}},
  \bibinfo{author}{\bibfnamefont{E.}~\bibnamefont{Tiemann}},
  \bibinfo{author}{\bibfnamefont{M.}~\bibnamefont{Lewenstein}},
  \bibinfo{author}{\bibfnamefont{S.}~\bibnamefont{Kotochigova}},
  \bibinfo{author}{\bibfnamefont{P.}~\bibnamefont{Julienne}}, \bibnamefont{and}
  \bibinfo{author}{\bibfnamefont{P.}~\bibnamefont{Zoller}},
  \bibinfo{journal}{Phys. Rev. Lett.} \textbf{\bibinfo{volume}{90}},
  \bibinfo{pages}{110401} (\bibinfo{year}{2003}).

\bibitem[{\citenamefont{Baranov et~al.}(2004)\citenamefont{Baranov, Dobrek, and
  Lewenstein}}]{Baranov04}
\bibinfo{author}{\bibfnamefont{M.}~\bibnamefont{Baranov}},
  \bibinfo{author}{\bibfnamefont{{\L}.}~\bibnamefont{Dobrek}},
  \bibnamefont{and}
  \bibinfo{author}{\bibfnamefont{M.}~\bibnamefont{Lewenstein}},
  \bibinfo{journal}{Phys. Rev. Lett.} \textbf{\bibinfo{volume}{92}},
  \bibinfo{pages}{250403} (\bibinfo{year}{2004}).

\bibitem[{\citenamefont{Pedri et~al.}(2008)\citenamefont{Pedri, Palo, Orignac,
  Citro, and Chiofalo}}]{Pedri08}
\bibinfo{author}{\bibfnamefont{P.}~\bibnamefont{Pedri}},
  \bibinfo{author}{\bibfnamefont{S.~D.} \bibnamefont{Palo}},
  \bibinfo{author}{\bibfnamefont{E.}~\bibnamefont{Orignac}},
  \bibinfo{author}{\bibfnamefont{R.}~\bibnamefont{Citro}}, \bibnamefont{and}
  \bibinfo{author}{\bibfnamefont{M.~L.} \bibnamefont{Chiofalo}},
  \bibinfo{journal}{Phys. Rev. A} \textbf{\bibinfo{volume}{77}},
  \bibinfo{pages}{015601} (\bibinfo{year}{2008}).

\bibitem[{\citenamefont{Marzok et~al.}(2007)\citenamefont{Marzok, Deh,
  Courteille, and Zimmermann}}]{Marzok07}
\bibinfo{author}{\bibfnamefont{C.}~\bibnamefont{Marzok}},
  \bibinfo{author}{\bibfnamefont{B.}~\bibnamefont{Deh}},
  \bibinfo{author}{\bibfnamefont{P.}~\bibnamefont{Courteille}},
  \bibnamefont{and}
  \bibinfo{author}{\bibfnamefont{C.}~\bibnamefont{Zimmermann}},
  \bibinfo{journal}{Phys. Rev. A} \textbf{\bibinfo{volume}{76}},
  \bibinfo{pages}{052704} (\bibinfo{year}{2007}).

\bibitem[{\citenamefont{Esposti and Werner}(1990)}]{esposti:3351}
\bibinfo{author}{\bibfnamefont{A.~D.} \bibnamefont{Esposti}} \bibnamefont{and}
  \bibinfo{author}{\bibfnamefont{H.-J.} \bibnamefont{Werner}},
  \bibinfo{journal}{J. Chem. Phys.} \textbf{\bibinfo{volume}{93}},
  \bibinfo{pages}{3351} (\bibinfo{year}{1990}).

\bibitem[{\citenamefont{Arndt et~al.}(1997)\citenamefont{Arndt, Dahan,
  Gu\'ery-Odelin, Reynolds, and Dalibard}}]{Arndt97}
\bibinfo{author}{\bibfnamefont{M.}~\bibnamefont{Arndt}},
  \bibinfo{author}{\bibfnamefont{M.~B.} \bibnamefont{Dahan}},
  \bibinfo{author}{\bibfnamefont{D.}~\bibnamefont{Gu\'ery-Odelin}},
  \bibinfo{author}{\bibfnamefont{M.}~\bibnamefont{Reynolds}}, \bibnamefont{and}
  \bibinfo{author}{\bibfnamefont{J.}~\bibnamefont{Dalibard}},
  \bibinfo{journal}{Phys. Rev. Lett.} \textbf{\bibinfo{volume}{79}},
  \bibinfo{pages}{625} (\bibinfo{year}{1997}).

\bibitem[{\citenamefont{Aubin et~al.}(2006)\citenamefont{Aubin, Myrskog,
  Extavour, Leblanc, McKay, Stummer, and Thywissen}}]{Aubin06}
\bibinfo{author}{\bibfnamefont{S.}~\bibnamefont{Aubin}},
  \bibinfo{author}{\bibfnamefont{S.}~\bibnamefont{Myrskog}},
  \bibinfo{author}{\bibfnamefont{M.~H.~T.} \bibnamefont{Extavour}},
  \bibinfo{author}{\bibfnamefont{L.~J.} \bibnamefont{Leblanc}},
  \bibinfo{author}{\bibfnamefont{D.}~\bibnamefont{McKay}},
  \bibinfo{author}{\bibfnamefont{A.}~\bibnamefont{Stummer}}, \bibnamefont{and}
  \bibinfo{author}{\bibfnamefont{J.~H.} \bibnamefont{Thywissen}},
  \bibinfo{journal}{Nat. Phys.} \textbf{\bibinfo{volume}{2}},
  \bibinfo{pages}{384} (\bibinfo{year}{2006}).

\bibitem[{\citenamefont{Treutlein et~al.}(2006)\citenamefont{Treutlein,
  H{\"a}nsch, Reichel, Negretti, Cirone, and Calarco}}]{Treutlein06}
\bibinfo{author}{\bibfnamefont{P.}~\bibnamefont{Treutlein}},
  \bibinfo{author}{\bibfnamefont{T.~W.} \bibnamefont{H{\"a}nsch}},
  \bibinfo{author}{\bibfnamefont{J.}~\bibnamefont{Reichel}},
  \bibinfo{author}{\bibfnamefont{A.}~\bibnamefont{Negretti}},
  \bibinfo{author}{\bibfnamefont{M.~A.} \bibnamefont{Cirone}},
  \bibnamefont{and} \bibinfo{author}{\bibfnamefont{T.}~\bibnamefont{Calarco}},
  \bibinfo{journal}{Phys. Rev. A.} \textbf{\bibinfo{volume}{74}},
  \bibinfo{pages}{022312} (\bibinfo{year}{2006}).

\end{thebibliography}

\end{document}